\documentclass[10pt, conference]{IEEEtran}

\pdfoutput=1

\usepackage{amsmath}
\usepackage{graphicx}
\usepackage{booktabs} 
\usepackage[font=small,font={small,sf},labelfont=bf]{caption}
\usepackage[labelformat=simple]{subcaption}

\usepackage{colortbl}
\usepackage{booktabs}
\usepackage{array}
\usepackage{multirow}
\usepackage[table,xcdraw]{xcolor}
\usepackage{hyperref}

\newcommand{\todo}[1]{\textcolor{red}{}}
\newcommand{\reduceV}{\vspace{-.7cm}} 
\newcommand{\reduceSV}{\vspace{-.2cm}}

\newcolumntype{P}[1]{>{\centering\arraybackslash}p{#1}}
\newcolumntype{M}[1]{>{\centering\arraybackslash}m{#1}}

\DeclareGraphicsExtensions{.pdf,.eps,.jpg}

\begin{document}
\title{Cloud-based or On-device: \\An Empirical Study of Mobile Deep Inference}

\author{
\IEEEauthorblockN{Tian Guo}
\IEEEauthorblockA{Computer Science Department \\
Worcester, MA, USA \\
Worcester Polytechnic Institute\\ 
\href{mailto:tian@cs.wpi.edu}{tian@cs.wpi.edu}}}

\maketitle

\begin{abstract}
Modern mobile applications are benefiting significantly from the advancement in deep
learning, 
e.g., implementing real-time image recognition and conversational system. 
Given a trained deep learning model, applications usually need to perform a
series of matrix operations based on the input data, in order to infer possible
output values. 
Because of computational complexity and size constraints, these trained models 
are often hosted in the cloud. 
To utilize these cloud-based models, mobile apps will have to send input data 
over the network. 
While cloud-based deep learning can provide reasonable response time
for mobile apps, it restricts the use case scenarios, e.g. mobile apps
need to have network access. With mobile specific deep learning optimizations, 
it is now possible to employ on-device inference. However, because mobile 
hardware, such as GPU and memory size, can be very limited when compared to its desktop counterpart, it is important to understand the feasibility of this new on-device 
deep learning inference architecture. 
In this paper, we empirically evaluate the inference performance of three Convolutional Neural Networks (CNNs) using a benchmark Android application we developed. 
Our measurement and analysis suggest that on-device inference can cost up to two orders of magnitude greater response time and energy when compared to cloud-based inference,
and that loading model and computing probability are two performance bottlenecks for on-device deep inferences.
 \end{abstract}
 

\IEEEpeerreviewmaketitle

\section{Introduction}
\label{sec:intro}

\begin{figure}[t]
\centering
\includegraphics[width=.4\textwidth]{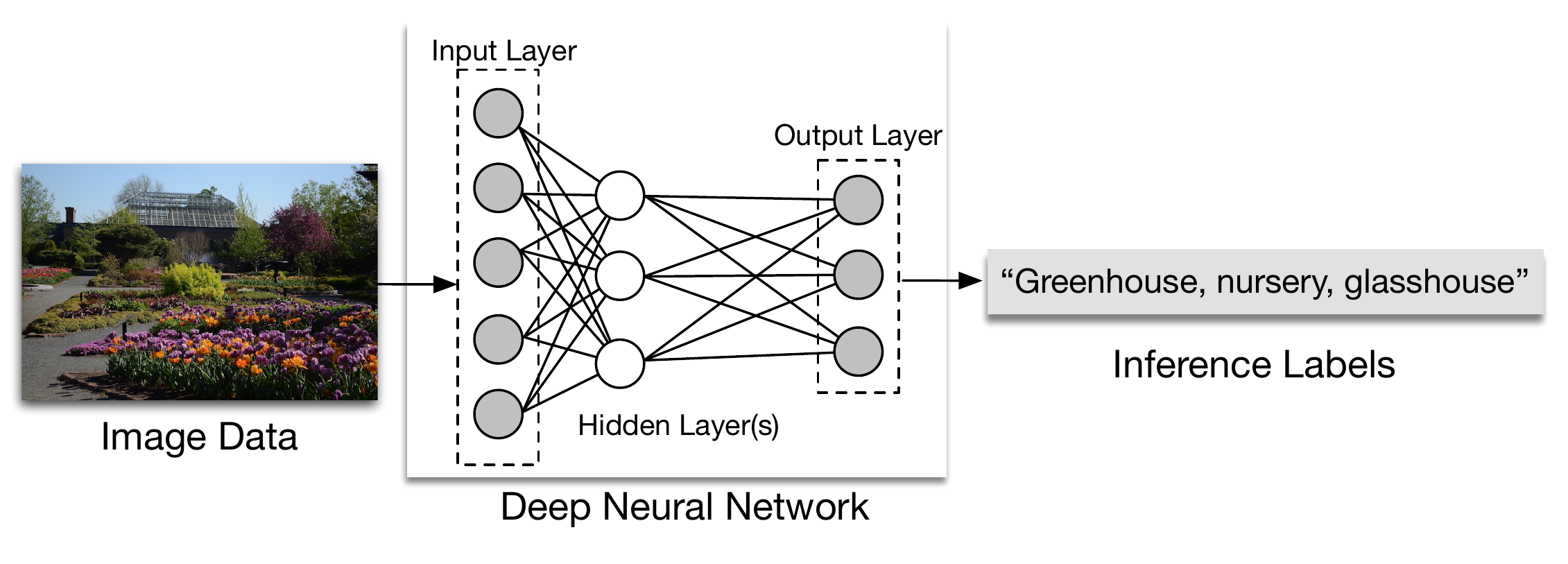}
\caption{
Object recognition with deep neural networks.
An image passes through a deep neural network that consists of layers of neurons. The output layer produces the inference labels that best describe the image.}
\reduceV
\label{fig:object_recognition}
\end{figure}

Deep learning has started to gain popularity in powering up modern mobile applications. Training deep neural networks requires access to large amounts of 
data and computing powers. As a result, these neural networks are often trained by leveraging cheaper, yet more powerful cloud GPU clusters. Once trained, the inference phase can be completed in a
reasonable amount of time, e.g., less than one second, using a single machine. Pre-trained models can be hosted for private use or offered as public cloud deep learning services~\cite{ws:googleCloudVision,ws:clarifai}. 
To utilize cloud-based pre-trained models,
mobile app developers use exposed cloud APIs to offload deep learning inference tasks, such as object recognition shown in Figure~\ref{fig:object_recognition}, to the hosting server. Mobile apps that execute inference tasks this way is referred to as \emph{cloud-based} deep inference. 

Despite their increasing popularity, the use case scenarios of cloud-based deep inference 
can be limited due to data privacy concern, unreliable network condition, and impact on battery life. 
Alternatively, we can perform inference tasks locally using mobile CPU and GPU~\cite{cnndroid:2016}. We refer to this mobile deep learning approach as \emph{on-device} deep inference. 
On-device deep inference can be a very attractive alternative to the cloud-based approach, e.g., by providing mobile applications the ability to function even without network access. 

Given the above two design choices for implementing deep inference, it is beneficial for developers to understand the performance differences. However, it is not straightforward 
to reason about mobile apps performance when using \emph{on-device} deep inference.  
The difficulties can be attributed to the following reasons. First, deep neural networks (DNN) can
differ vastly in terms of network architecture, number of parameters, and model sizes (see Figure~\ref{fig:dnn_models}). Second, the inference tasks can be of different complexities depending on the input data, e.g., large images vs. small images, and the DNN model in use. Third, mobile devices often have heterogenous resource capacities and can exhibit different runtime behaviors, such as garbage collection activities, given different deep learning models and inference tasks combinations.

To address the challenges of understanding deep learning inference, in this paper we develop a mobile application benchmark 
that allows end users to supply configurations including inference mode, model, and input data. 
We conduct a detailed measurement study using our mobile application with cloud-based and on-device deep inference,
three convolutional neural networks, and a dataset of fifteen images. Our evaluation shows that cloud-based approach can save up 
to two orders of magnitude in terms of both end-to-end response time and mobile energy consumption. Further, we analyze the performance differences 
between on-device and cloud-based approaches with an in-depth analysis of performance bottlenecks. Our analysis suggests 
that loading and computing using deep learning models dominate the end-to-end inference time. 
Lastly, we find that current on-device approach, when utilizing mobile GPU, provides reasonable average inference time of 2.2 seconds if models are preloaded into the memory.

\begin{figure*}[t]
\centering
\begin{subfigure}{0.4\textwidth}
\includegraphics[width=\columnwidth]{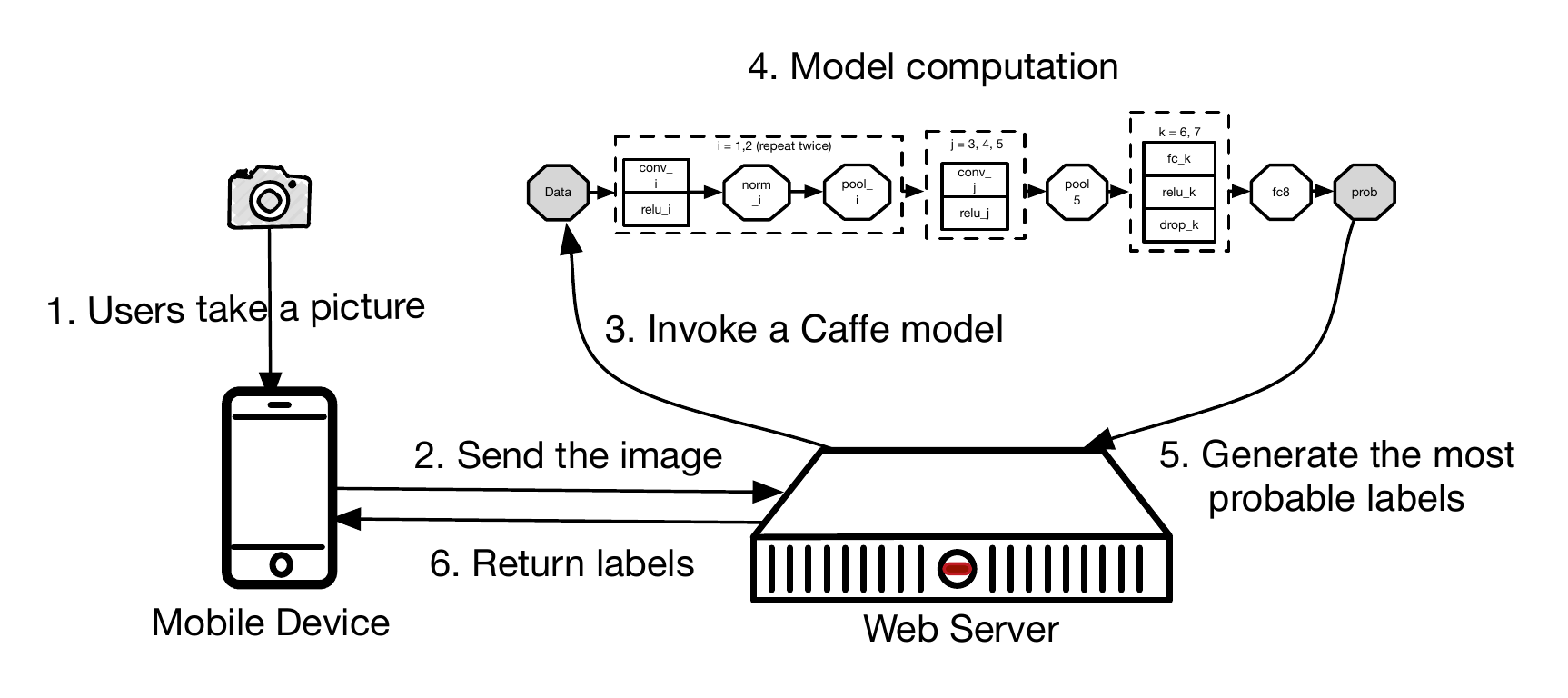}
\reduceV
\caption{\textnormal{Cloud-based inference.}}
\label{subfig:alexnet} 
\end{subfigure}
\qquad
\begin{subfigure}{.36\textwidth}
\includegraphics[width=\columnwidth]{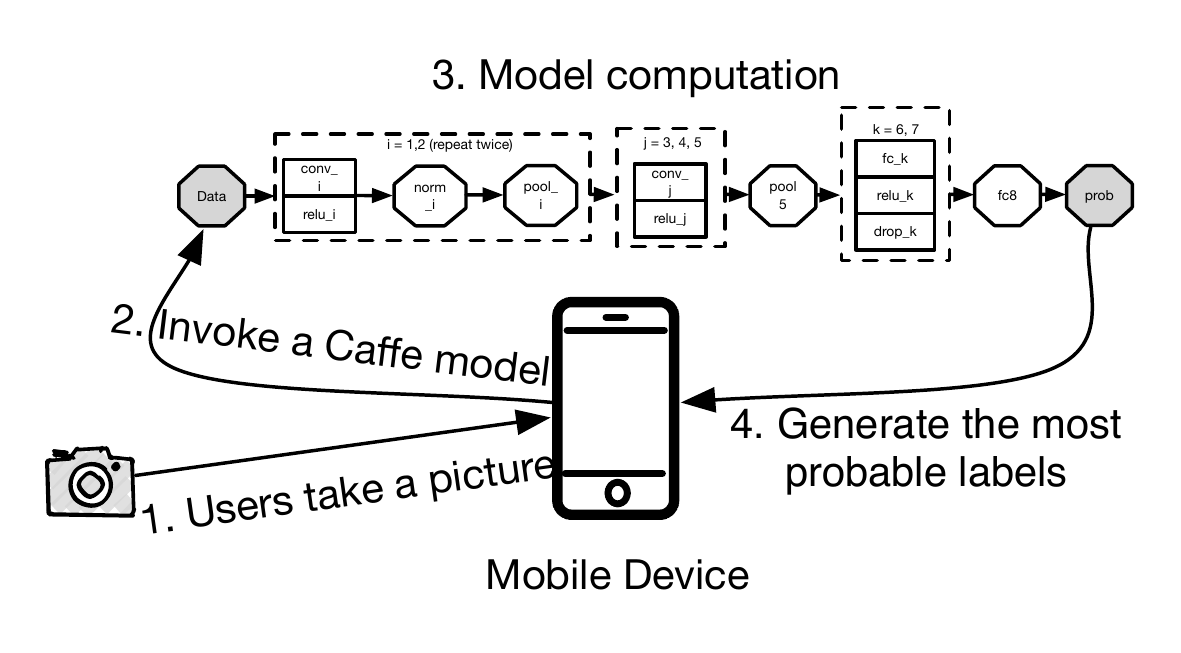}
\reduceV
\reduceSV
\caption{\textnormal{On-device inference.}}
\label{subfig:nin} 
\end{subfigure}
\reduceSV
\caption{
Design choices of deep learning powered mobile applications.
We use our implemented object recognition Android App as an example to illustrate the steps involved to perform cloud-based inference and on-device inference.}
\label{fig:exp_setup}
\end{figure*}

 \section{Background} 
\label{sec:back}

In this section, we first provide a background on deep learning models 
and platforms. We then discuss two design choices for implementing deep learning 
powered mobile apps. Lastly, we provide a brief explanation of mobile OS memory resource 
management and its associated performance implication.

\subsection{Deep Learning Models}

Deep learning refers to a class of artificial neural networks (ANNs) for learning the right representation of input data
and is widely used for visual and speech recognition~\cite{Goodfellow-et-al-2016}. In this paper, we focus on a specific visual recognition task called object recognition that maps 
an image to a list of most probable text labels using deep neural networks (DNNs), as illustrated in Figure~\ref{fig:object_recognition}. DNNs usually consist of one input layer and one output layer, and a number of hidden layers. 
Each layer can consist of different numbers of processing units (neurons). Given an image, DNNs use it as input to 
initialize the first layer, pass it through processing units in hidden layers, and eventually generate a probability distribution over label categories
at the final layer. The inference labels correspond to the categories with the highest probability values. The complexity of the inference tasks depends on the computation defined in each unit as well as the total number of units and layers. 

In this paper, we focus on a special class of deep learning models called convolutional neural networks (CNNs)~\cite{alexnet:2012,nin:2014,SqueezeNet:2016}. 
CNNs are widely used for visual recognition tasks due to their high accuracy. CNNs usually consist of many different types of 
layers, such as convolutional layers, pooling layers, and fully-connected layers (refer to Figure~\ref{fig:dnn_models}). Each layer takes data from 
previous layer and apply predefined computation in parallel. For example, convolutional layers use both linear convolutional filters and 
nonlinear activation functions to generate feature maps~\cite{lecun-01a}, while pooling layers control overfitting by reducing parameters in the representation~\cite{ws:cs231n}.

There are a number of popular deep learning frameworks, 
such as Caffe~\cite{caffe}, Torch~\cite{collobert2002torch}, TensorFlow~\cite{tensorflow2015-whitepaper}, 
that ease the training and deploying of deep learning models.
Different frameworks require different syntaxes to describe CNNs and have different trade-offs
for training and inference~\cite{bahrampour2015comparative} phases. Because Caffe provides a flexible way to define CNN layers and has a large collection of pre-trained models~\cite{ws:caffeModelZoo},  
we choose to evaluate CNN models trained using the Caffe framework in this paper. A pre-trained Caffe model contains a binary \texttt{caffemodel} file that describes model parameters, a \texttt{prototxt} file that describes model network,
and an accompanying text file of labels. 

\subsection{Mobile Apps} 

\subsubsection{Cloud-based vs. on-device Inference}
Deep learning powered mobile apps can be roughly categorized into two types (as shown in Figure~\ref{fig:exp_setup}) : the cloud-based inference and on-device inference. 
The key differences between these two architectures are where the CNN models are stored, and how the inference task is executed. 
Cloud-based inference leverages pre-trained models in powerful cloud servers and is by far the more popular design choice when it comes to designing mobile deep learning apps.

Alternatively, mobile apps can be built upon on-device inference. In essence, this means CNN models will be stored on mobile device 
and inference tasks will be executed using mobile CPU or GPU. Although conceptually simple, it is not always straightforward to deploy 
CNN models into mobile devices. For example, existing models that are designed to run on powerful servers from the outset can contain hundreds of layers and millions of parameters, therefore are not suitable
to run on resource-constrained mobile devices.
\footnote{Mobile-specific platforms~\cite{ws:tensorflowlite,ws:caffe2go} and mobile-specific models~\cite{ws:mobilenets}, designed by companies such as Google and Facebook, can be a promising approach towards efficient on-device inference for new mobile hardwares.}

In this paper, we look at two existing approaches that enable on-device inference. The first approach relies on porting existing frameworks~\cite{ws:torchmobile,ws:caffeMobile} to mobile platforms so that CNN models can run on the mobile platform.  
When developing Android apps using Caffe Android Lib~\cite{ws:caffeMobile}, app developers first need to compile 
different versions of \texttt{libcaffe.so} and \texttt{libcaffe\_jni.so} for all supported mobile CPU and instruction sets. These compiled library files then need to be loaded before applications perform inference tasks. However, because mobile GPUs are very different from desktop GPUs, currently none of the ported mobile libraries support executing model computation using mobile GPUs. The second approach relies on third-party libraries that convert existing models to supported formats~\cite{cnndroid:2016} to take advantage of the mobile GPU. For example, CNNDroid expresses CNN layers in \texttt{RenderScript} kernels so that 
the RenderScript runtime framework can parallelize model computations across both CPUs and GPUs~\cite{ws:renderscript}.

\subsubsection{Lifecycle Management and Its Performance Implication}
When an user first launches the mobile app, Android OS will first call the \texttt{onCreate()} method inside the launcher main activity. 
After successfully setting up and initializing states, the activity runs in the foreground of the screen. 
This running activity is called foreground activity and at any given time, there is only one such activity. 
Android Runtime (ART) automatically manages application 
memory by 
using concurrent mark sweep (CMS)  garbage collection algorithm that 
is optimized for interactive applications. By default, Android OS allocates a heap size indicated by \texttt{getMemoryClass()} method. 
For memory-intensive apps, such as CNN based mobile apps, we can request to run the app with large heap. By doing so, Android OS might allocate a heap size indicated by \texttt{getLargeMemoryClass()} method. 
For example, in Nexus 5, we can increase the application heap from the default 192 MB to 512 MB with large heap option turned on. However, for memory constrained 
mobile devices,  Android OS might still allocate the default heap size to applications. 
When an activity is in the background, it 
can be killed by Android OS when memory is needed elsewhere, 
e.g. by another running activity~\cite{ws:androidActivity}. Since a CNN model can be up to hundreds of MBs, and therefore apps of on-device inference architecture are more likely to be killed by OS in a memory-constrained mobile device to free up memory.  When users need to interact with these killed apps, the app will have to be completely restarted and restored to its previous state, incurring undesirable startup latency.

 \begin{table}[t]
\centering
\caption{Hardware Specification of mobile device and cloud server used in our evaluation.}
\label{tbl:hardware}
\resizebox{\columnwidth}{!}{
\begin{tabular}{l|ccccc}
\rowcolor[HTML]{EFEFEF} 
\begin{tabular}[c]{@{}l@{}}Inference \\ Engine\end{tabular} & \textbf{CPU}                                                                                & \textbf{GPU}                                                                         & \textbf{Mem.} & \textbf{Storage} & \textbf{Battery} \\ \hline
Nexus 5                                                     & \begin{tabular}[c]{@{}c@{}}2.26 GHz \\ quad-core \\ (Krait 400)\end{tabular}          & \begin{tabular}[c]{@{}c@{}}Adreno 330 \\ (129.8 GFLOPS)\end{tabular}        & 2GB  & 16GB   & \begin{tabular}[c]{@{}c@{}}2300 mAh\\/8.74Wh\end{tabular}                                           \\ \hline
g2.2xlarge                                                  & \begin{tabular}[c]{@{}c@{}}2.6 GHz \\ eight-core \\ (Intel Xeon E5-2670)\end{tabular} & \begin{tabular}[c]{@{}c@{}}NVIDIA \\ GRID K520\\ (2448.4 GFLOPS)\end{tabular} & 15GB & 60GB   & N/A                                                
\end{tabular}}
\end{table}

\begin{table}[t]
  \centering
        \caption{
        Summary of image data sets. Each row describes the image dimension in terms of width by height, and the image size in KB. \todo{remove}}
      \label{tbl:images}
       \resizebox{\columnwidth}{!}{
     \begin{tabular}{M{1cm}M{2cm}M{2cm}M{2cm}M{2cm}M{2cm}M{2cm}}
     \toprule
     \textbf{Dataset} & \textbf{Dimension(WxH)/ 
     Size(KB)} & \raisebox{-\totalheight}{\includegraphics[width=15mm, height=15mm]{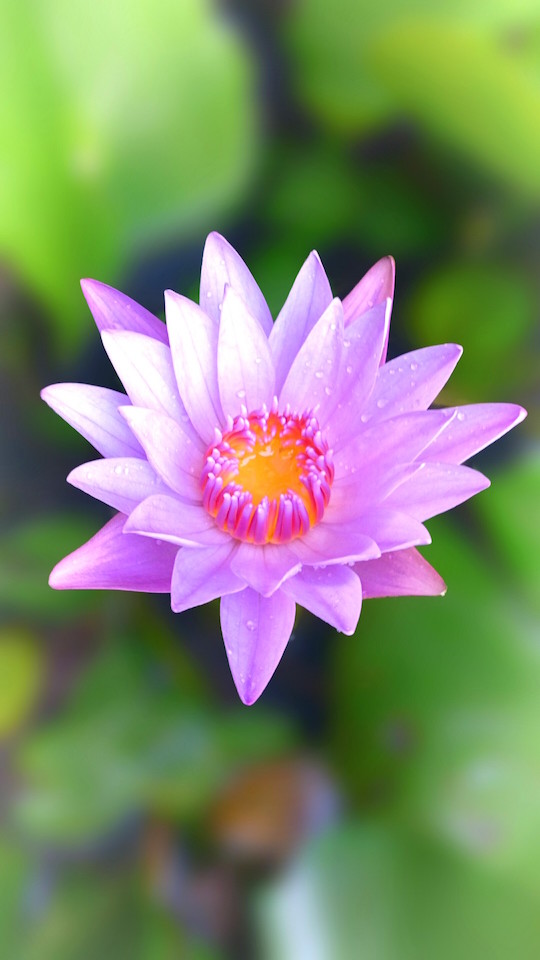}}
      & 
           \raisebox{-\totalheight}{\includegraphics[width=15mm, height=15mm]{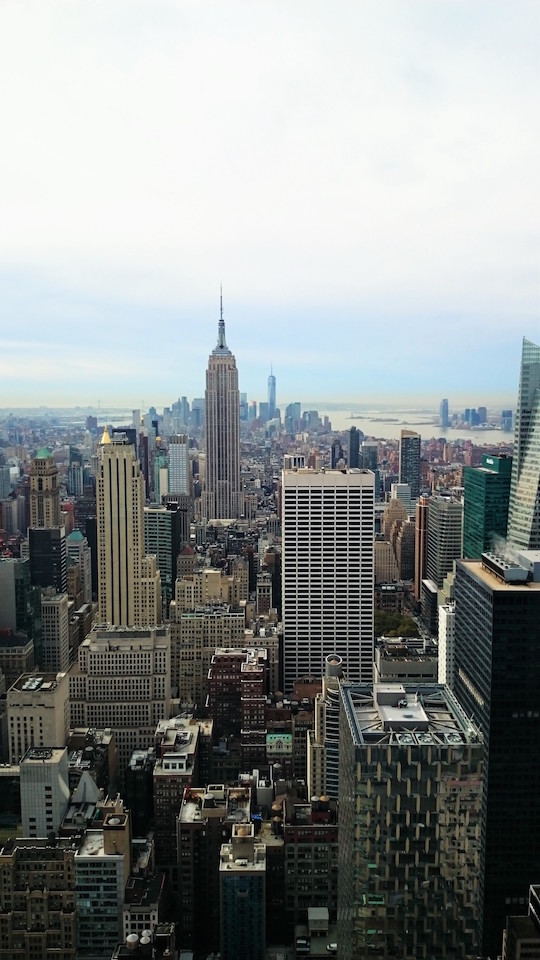}}
      & 
	           \raisebox{-\totalheight}{\includegraphics[width=15mm, height=15mm]{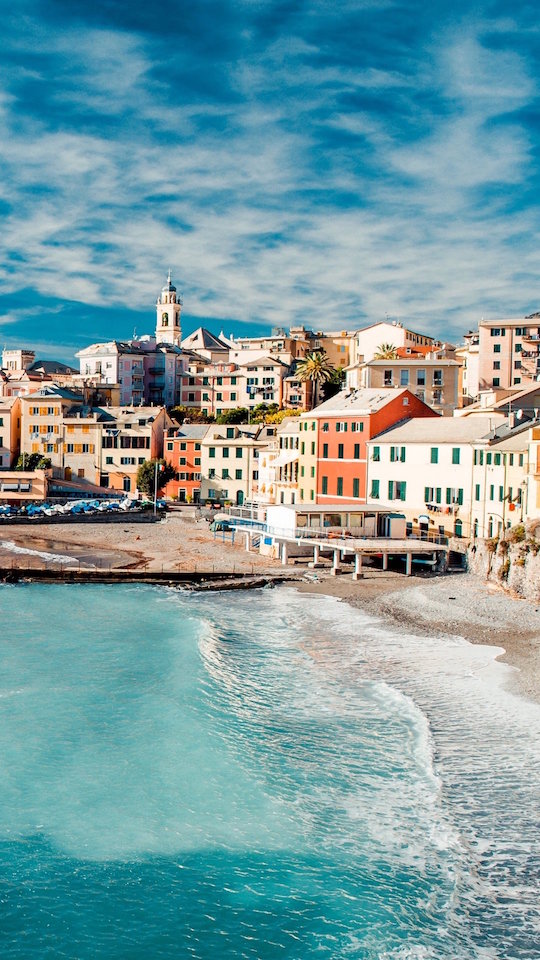}}
      & 
      	           \raisebox{-\totalheight}{\includegraphics[width=15mm, height=15mm]{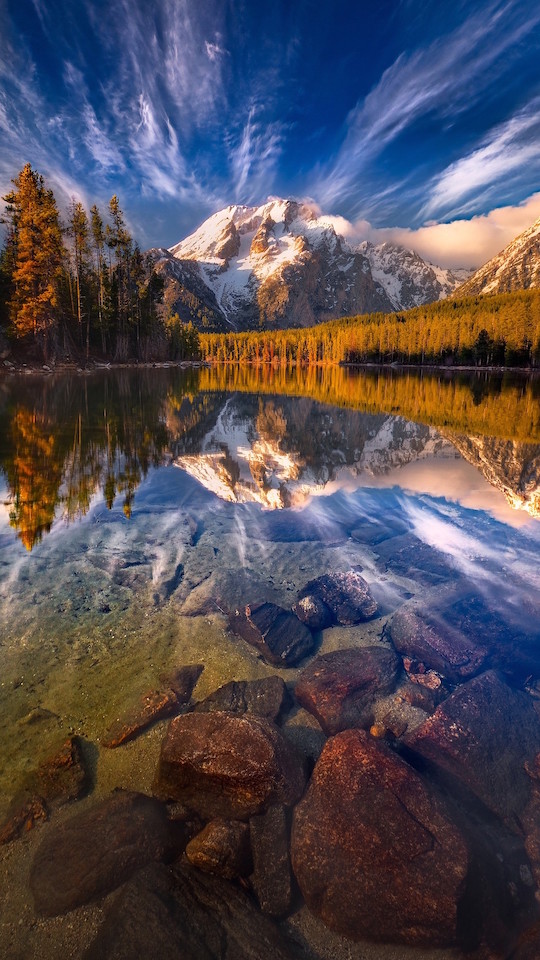}}
	& 
		           \raisebox{-\totalheight}{\includegraphics[width=15mm, height=15mm]{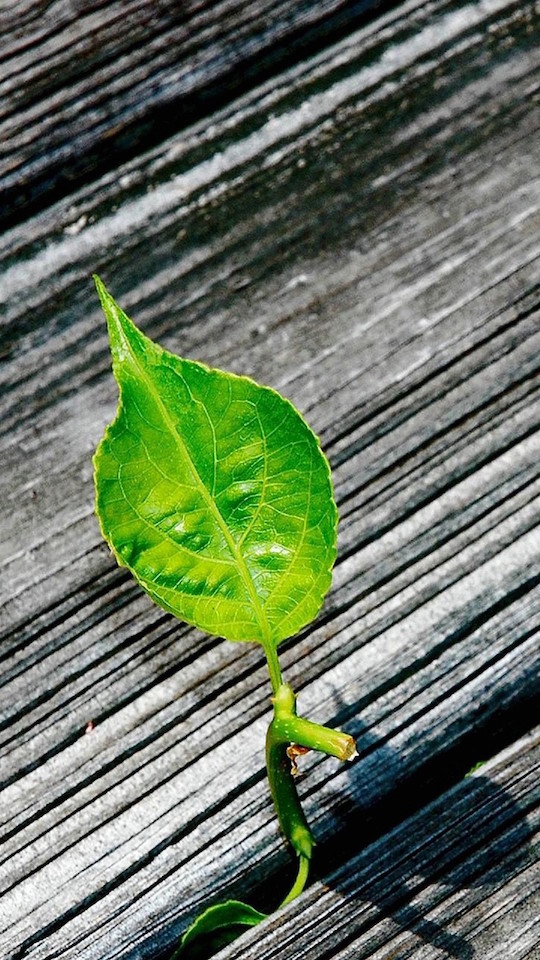}}
      \\ \hline \\
      1 &2160 x 3840 & 785 & 1362  & 1396 & 1528 & 2599 \\ 
      2&1080 x 1920 & 251 & 479 & 547 & 739 & 576 \\ 
      3 &540 x 960 & 82 &  146 &  165 &  196 &  226 	\\ 
      \bottomrule
      \end{tabular}}
 \end{table}

\section{Mobile Benchmark implementation} 

\label{sec:impl}
We implement an object recognition Android app that 
supports both cloud-based and on-device inference. 
As shown in Figure~\ref{fig:exp_setup}, our Android app 
takes images as data input, invokes one of 
the CNN models, 
generates a probability distribution over labels, 
and displays the top five most probable labels to the user.

In the cloud-based inference mode, image data needs to be sent from our 
mobile app to an Apache web server hosted inside the Amazon Virginia data center. 
To achieve this, our mobile app specifies both the IP address of the cloud server and
the \texttt{php} script name when creating the HTTP 
connection. After the HTTP connection is successfully established, 
our mobile app will send the original or downscaled image of 224 by 224 pixels in 
dimension, to the web server. This downsizing step is because all three deep learning 
models we are using only require bitmaps of the scaled dimension. 
After the image finishes uploading, the web server will then invoke the specified Caffe 
model, use either CPU-only or GPU accelerated Caffe framework for probability computation, and return the top five labels to the mobile app. 

In the on-device inference mode, both the selected CNN model and 
the image bitmap object are loaded into mobile device's memory. Depending 
on which framework is selected, our mobile app will 
use either CPU-only Caffe Library or GPU-enabled CNNDroid to generate the probability distribution over labels.

\section{Mobile Deep Inference Evaluation}
\subsection{Experimental Setup}

\begin{figure*}[t]
\centering
\begin{subfigure}{0.4\textwidth}
\includegraphics[width=\columnwidth]{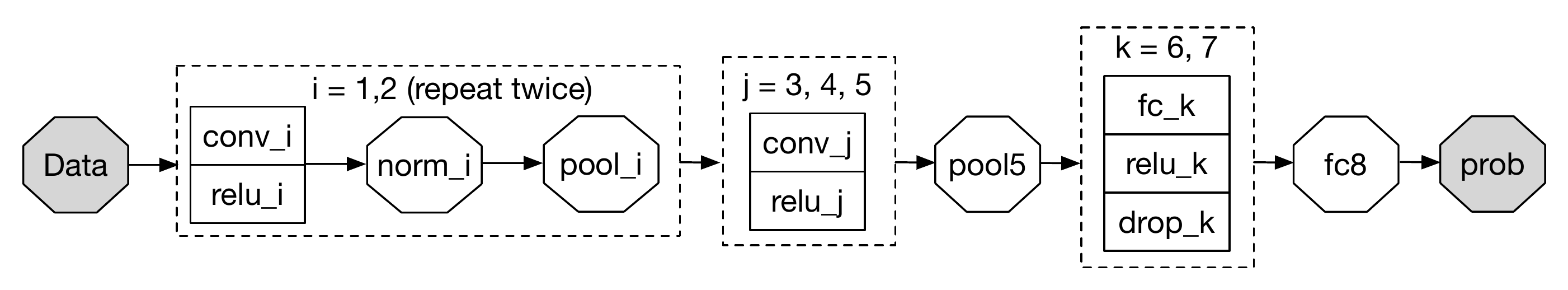}
\caption{\textnormal{AlexNet model.}}
\label{subfig:alexnet} 
\end{subfigure}
\hfill
\begin{subfigure}{.55\textwidth}
\includegraphics[width=\columnwidth]{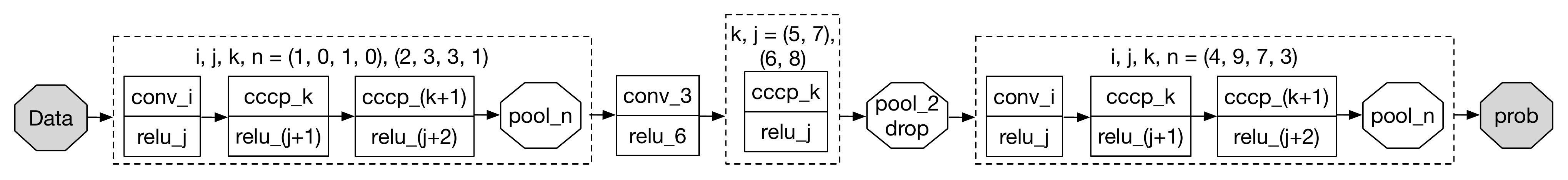}
\caption{\textnormal{NIN model.}}
\label{subfig:nin} 
\end{subfigure}
\begin{subfigure}{\textwidth}
\includegraphics[width=\columnwidth]{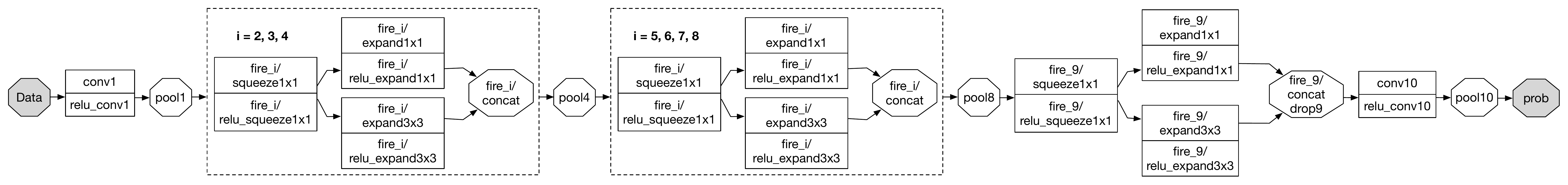}
\caption{\textnormal{SqueezeNet model.}}
\label{subfig:squeeze} 
\end{subfigure}
\caption{
Neural Network architecture visualization.
For each neural network, we show a simplified version based on a web tool~\cite{netscope}. These three convolution neural networks achieve very close top-5 ImageNet accuracy of around 80\%~\cite{SqueezeNet:2016,imagenet:accuracy}, but have vastly different model sizes. AlexNet~\cite{alexnet:2012} is 233 MB, NIN~\cite{nin:2014} is 29 MB, and SqueezeNet~\cite{SqueezeNet:2016} is 4.8 MB.}
\label{fig:dnn_models}
\end{figure*}

Our evaluation consists of empirical measurement of performing inference tasks using both cloud-based and on-device deep learning models,
as illustrated in Figure~\ref{fig:exp_setup}. 
We use our implemented mobile deep inference benchmark application, described in Section~\ref{sec:impl}, to perform both inference modes. 
For evaluating cloud-based setup, we first install Apache web server on a cloud server, and then store three CNN models in the web server. 
We select the cheapest GPU server, i.e., \texttt{g2.2xlarge}, from Amazon Virginia data center for both the CPU and GPU cloud-based inference. 
All our cloud servers run Ubuntu 14.04. 
For evaluating on-device inference, we use a LG Nexus 5 (late 2013)  
mobile phone that runs Android 6.0.1 Marshmallow 
and is on university campus Wi-Fi. 
Hardware specifications can be found in Table~\ref{tbl:hardware}.

We selected three Caffe-based CNN models: AlexNet~\cite{alexnet:2012}, NIN~\cite{nin:2014}, and SqueezeNet~\cite{SqueezeNet:2016},
as shown in Figure~\ref{fig:dnn_models}.
We choose to evaluate our mobile application using these models 
because they provide very similar top-5 error rates\footnote{Top-5 error rate is defined as the percentage of incorrectly labeled test images in the top five most probable labels.} 
on ImageNet data set, while differ vastly in terms of model sizes. All three models only require images of dimension 224 by 224 pixels. 
Our inference input data set contains three groups of a total fifteen images, as summarized in Table~\ref{tbl:images}. Each group consists of images of the same dimension. 
The second and third groups are generated by resizing the original images in the first group by 
scale factors of two and four respectively.

To measure the time taken to perform each step in an inference 
task, we instrumented our Android app to output event timestamps
to a log file. We use \texttt{Logcat}, a command-line tool, to pull 
log files of each experiment run from the Nexus 5 through 
a laptop running within the same university campus network over Wi-Fi.
To prevent inaccurate power measurement, we avoided connecting 
the Nexus 5 to laptop through USB in our experiments. We also ensure the mobile device is fully charged at 100\% at the start of every measurement session.
\texttt{Logcat} logs contains system events, including Android Runtime (ART) 
garbage collection logs, and application-level logs. 

To measure the power consumption and resource utilization of our mobile app, we use 
the \texttt{Trepn} profiler~\cite{ws:trepn} and save the profile results in \texttt{csv} format 
for offline analysis. 
We follow the best practices to reduce a profiler's impact on measurement inaccuracy 
by configuring \texttt{Trepn} to sample every 100 ms for 
three data points of interests: battery consumption, 
normalized CPU usage, 
and the GPU load. 

For each experiment run, the user first launches the \texttt{Trepn} 
profiler mobile app. Inside the profiler app, the user can select 
to profile our developed object recognition app. 
This will automatically launch our app, and present the UI for setting up
experiment configuration.
The user can select either to run the experiment in cloud or mobile 
modes, indicate the deep learning models to use, and choose 
the dataset to perform the object recognition on.

\begin{table*}[]
\centering
\caption{
Comparisons of cloud-based and on-device deep inference.
Cloud-based models outperform on-device models by at least six times. The model loading time dominates the object recognition task, and the time taken to compute the label probabilities dominates the inference time for on-device mode. In addition, on-device mode consumes twice as much battery, therefore consumes at least 12 times as much mobile energy. \todo{Define the recognition time, and the inference time.} }
\label{tbl:end2end}
\begin{tabular}{r|rrrr|rrrr}
\hline
\multicolumn{1}{c|}{}                                          & \multicolumn{4}{c|}{\cellcolor[HTML]{EFEFEF}\textbf{Object Recognition Time Breakdown {[}ms{]}}}                                                                                                                                                                                                                           & \multicolumn{4}{c}{\cellcolor[HTML]{EFEFEF}\textbf{On-device Resource Consumption}}                                                          \\ \cline{2-9} 
\multicolumn{1}{c|}{\multirow{-2}{*}{\textbf{Inference Mode}}} & \multicolumn{1}{c}{\begin{tabular}[c]{@{}c@{}}Load\\ model\end{tabular}} & \multicolumn{1}{c}{\begin{tabular}[c]{@{}c@{}}Rescale\\ bitmap\end{tabular}} & \multicolumn{1}{c}{\begin{tabular}[c]{@{}c@{}}Upload\\ bitmap\end{tabular}} & \multicolumn{1}{c|}{\begin{tabular}[c]{@{}c@{}}Compute\\ probability\end{tabular}} & \multicolumn{1}{c}{CPU {[}\%{]}} & \multicolumn{1}{c}{GPU {[}\%{]}} & \multicolumn{1}{c}{battery {[}mW{]}} & \multicolumn{1}{c}{Mem{[}MB{]}} \\ \hline
\cellcolor[HTML]{EFEFEF}Cloud+CPU                              & 0.00                                                                     & 76.16                                                                        & 36.83                                                                       & 238.60                                                                             & 6.22                             & 0.89                             & 1561.63                              & 1279.10                         \\ 
\cellcolor[HTML]{EFEFEF}Cloud+GPU                              & 0.00                                                                     & 76.16                                                                        & 36.83                                                                       & 18.60                                                                              & 6.36                             & 0.43                             & 1560.17                              & 1311.38                         \\
\cellcolor[HTML]{EFEFEF}Device+Caffe                           & 2422.13                                                                  & 79.98                                                                        & 0.00                                                                        & 8910.64                                                                            & 35.01                            & 0.14                             & 3249.01                              & 1637.16                         \\
\cellcolor[HTML]{EFEFEF}Device+CNNDroid                        & 61256.17                                                                 & 70.43                                                                        & 0.00                                                                        & 2131.70                                                                            & 22.20                            & 27.14                            & 2962.58                              & 1752.45                         \\
\hline
\end{tabular}
\end{table*}

\subsection{End-to-end Comparisons} 

In this section, we present the measurement results of object recognition time and mobile resource utilization for both cloud-based and on-device inference. 
\subsubsection{Object Recognition Time}
Table~\ref{tbl:end2end} summarizes the average end-to-end performance and resource consumption of executing object recognition using both cloud-based and on-device inference modes.
The task of object recognition is further broken down into four steps: loading CNN models into memory, downscaling image input to desired dimension, uploading input data to the 
cloud server, and computing the probability matrix. For each inference mode, we repeat the recognition tasks using all fifteen images and three CNN models. We measure the time to execute each step and calculate the average. We use CPU-only Caffe framework and GPU accelerated Caffe framework for toggling the CPU and GPU mode in our \texttt{g2.2xlarge} server. 
Note, the time to load models is negligible in the cloud-based scenario because models already reside inside the memory and can be used to execute the inference task immediately. Similarly, on-device mode does not incur any time for uploading image bitmaps. 
Recall, the inference time is the sum of rescaling, uploading bitmap and computing probability over one bitmap, and the recognition time is the sum of amortized model loading time over a batch of images and inference time. The average cloud-based inference time is 351.59 ms/131.59 ms when using CPU-only/GPU of a well-provisioned cloud instance hosted in a nearby data center. 
As shown, because inference tasks are typically data-parallel and therefore can be accelerated by up to 10x when using GPU. 
However, we should note that such results represent a lower bound performance of real-world setting. 
In a real-world deployment scenario, object recognition time can last much longer due to reasons such as overloaded cloud servers and variable mobile network conditions. 
The total inference time when running on-device is almost 9 seconds when using Caffe-based model, and 2.2 seconds when using CNNDroid model.

\subsubsection{Energy and Resource Consumption} 
Table~\ref{tbl:end2end} shows the resource consumption of mobile device when running the object recognition mobile application. 
As a baseline, we measure the performance when the device is idle and the screen is turned on. The CPU utilization and power consumption is 3.17\% and 1081.24 mW respectively. Cloud-based mode consumes roughly the same amount of CPU and 
44.4\% more power consumption comparing to the baseline. However, on-device mode 
not only incurs significantly higher CPU utilization (and in the case of CNNDroid, GPU utilization as well), but also require two times more power consumption when compared to the baseline. 
In all, we can calculate the energy consumption of different inference modes by multiplying the average inference (recognition) time by the average power consumption. Cloud-based inference requires as low as  
0.057 mWh energy when using faster GPU computation, and on-device based inference consumes up to 
8.11 mWh.

\emph{\textbf{Result:} Cloud-based inference exhibits substantial benefits in terms of inference response time and mobile energy savings over on-device inference, in this case by two orders of magnitude. This is due to more powerful processing power and shorter durations of inference. \todo{should we add to our policies?}}

\begin{figure*}[t]
\centering
\begin{subfigure}{0.28\textwidth}
\includegraphics[width=\columnwidth]{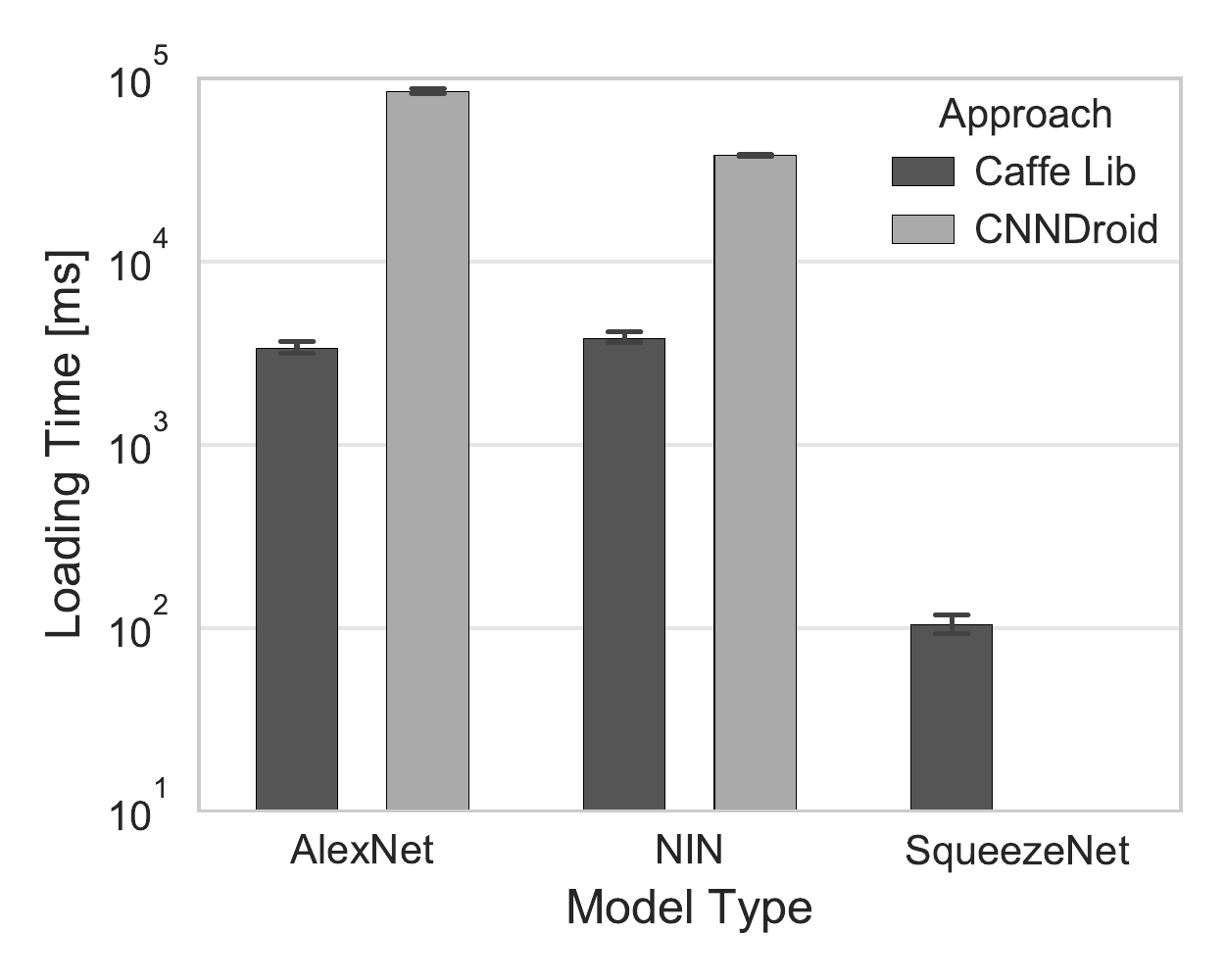}
\caption{\textnormal{Model loading time.}}
\label{subfig:model_loading_time} 
\end{subfigure}
\hfill
\begin{subfigure}{0.35\textwidth}
\includegraphics[width=\columnwidth]{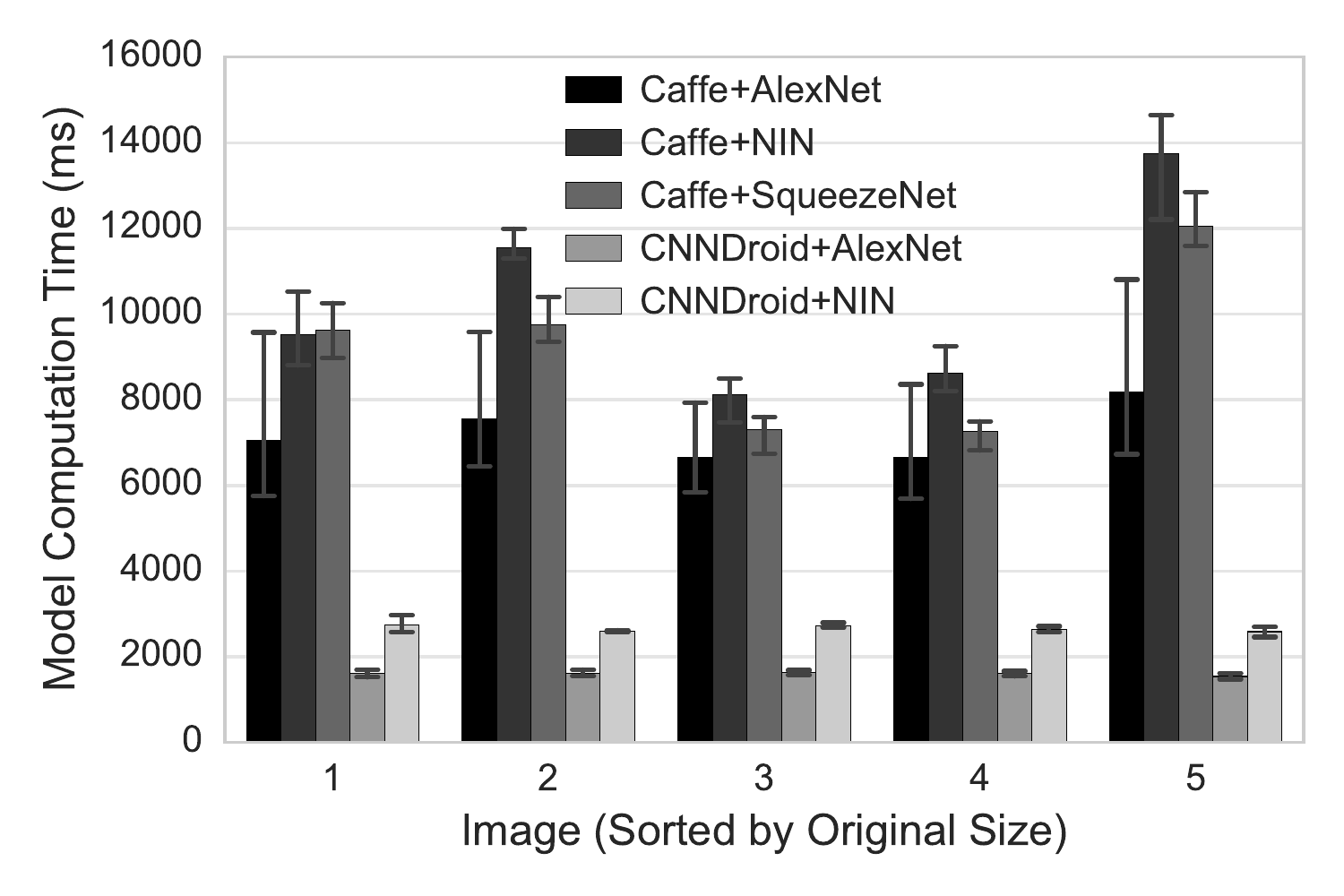}
\caption{\textnormal{Model computation time.}}
\label{subfig:model_computation_time} 
\end{subfigure}
\hfill
\begin{subfigure}{0.35\textwidth}
\includegraphics[width=\columnwidth]{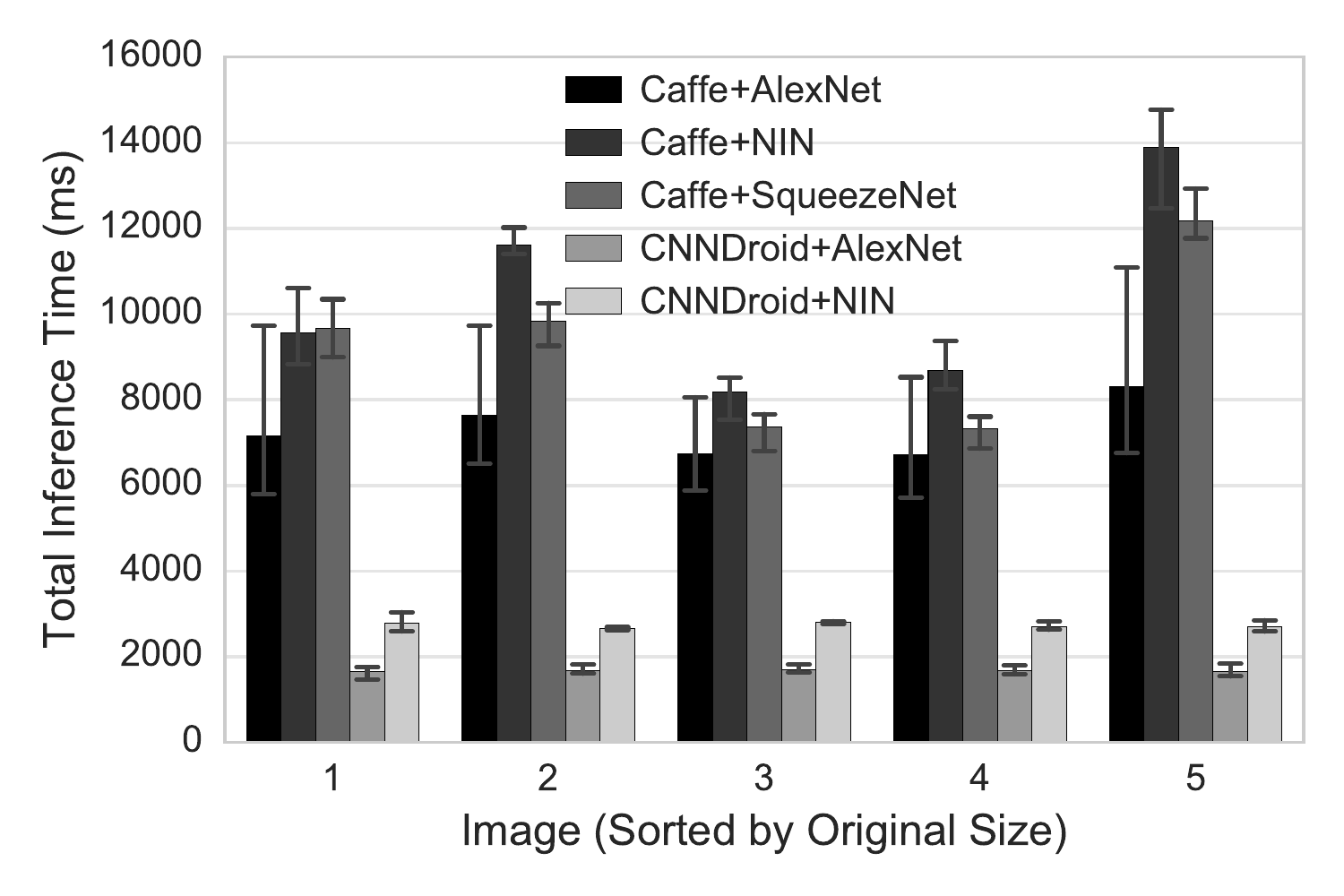}
\caption{\textnormal{Total inference time.}}
\label{subfig:model_inference_time} 
\end{subfigure}
\reduceSV
\caption{
On-device Inference Time.
We compare two approaches for device-based inference using three CNN models. When using CNNDroid-based approach, trained models need to be converted to supported format.}
\label{fig:dnn_models}
\end{figure*}

\begin{table}[]
\centering
\caption{
Summary of GC activities when using CNNDroid-based device inference.
ART uses the default CMS GC, and the GC time takes up to 9.89\% 
during model loading, and up to 25\% during user interactions. The average GC pause time can be 
up to 39.23 ms.}
\label{tbl:gc_impact}
\resizebox{\columnwidth}{!}{
\begin{tabular}{r|c|rrrr}
\hline
\multicolumn{1}{c|}{\begin{tabular}[c]{@{}c@{}}On-device \\ Inference\end{tabular}} & \cellcolor[HTML]{EFEFEF}\textbf{Phase}                                        & \multicolumn{1}{c}{\cellcolor[HTML]{EFEFEF}\textbf{Duration{[}ms{]}}} & \multicolumn{1}{c}{\cellcolor[HTML]{EFEFEF}\textbf{\begin{tabular}[c]{@{}c@{}}Num. \\ of GC\end{tabular}}} & \multicolumn{1}{c}{\cellcolor[HTML]{EFEFEF}\textbf{\begin{tabular}[c]{@{}c@{}}GC \\ Time {[}ms{]}\end{tabular}}} & \multicolumn{1}{c}{\cellcolor[HTML]{EFEFEF}\textbf{\begin{tabular}[c]{@{}c@{}}GC \\ Pause {[}ms{]}\end{tabular}}} \\ \hline
\cellcolor[HTML]{EFEFEF}CNNDroid+AlexNet                                            &                                                                               & 84537.33                                                              & 4.33                                                                                                       & 513.55                                                                                                           & 10.42                                                                                                             \\
\cellcolor[HTML]{EFEFEF}CNNDroid+NIN                                                & \multirow{-2}{*}{\begin{tabular}[c]{@{}c@{}}Load \\ Model\end{tabular}}       & 37975                                                                 & 16.67                                                                                                      & 3757.30                                                                                                          & 175.76                                                                                                            \\ \hline
\cellcolor[HTML]{EFEFEF}CNNDroid+AlexNet                                            &                                                                               & 11800                                                                 & 4                                                                                                          & 536.55                                                                                                           & 4.60                                                                                                              \\
\cellcolor[HTML]{EFEFEF}CNNDroid+NIN                                                & \multirow{-2}{*}{\begin{tabular}[c]{@{}c@{}}User \\ Interaction\end{tabular}} & 17166.67                                                              & 7                                                                                                          & 4307.18                                                                                                          & 274.66                                                                                                            \\ \hline
\end{tabular}}
\end{table}

\begin{figure}[t]
\centering
    \includegraphics[width=.4\textwidth]{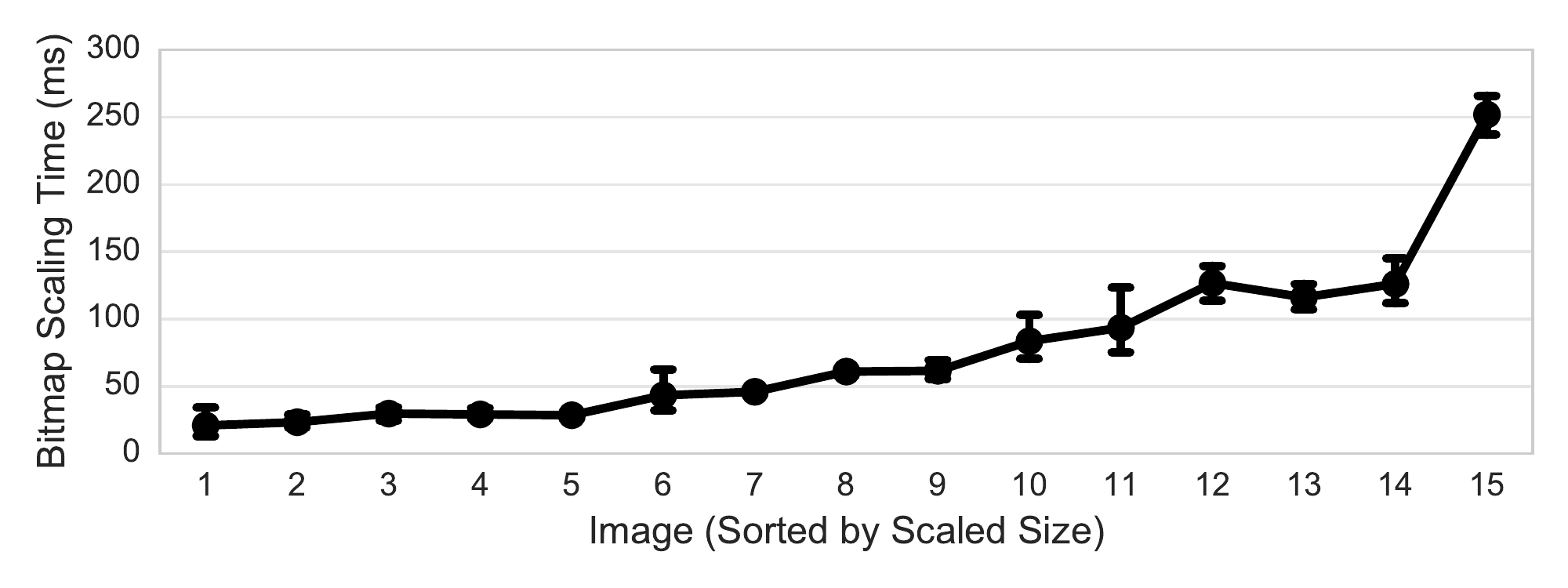}
    \caption{
    Bitmap downscaling time.
    The time taken to downscale the image in the mobile device grow with the image size. Specially, larger images also experiences proportionally longer scaling time because of the limited memory resources assigned to the mobile application. \todo{why not have x-axis as image size, probably because the datapoint is going to be clustered?}}
    \label{fig:scaling_time}
\end{figure}

\subsection{On-device Deep Inference Performance Analysis}
\label{subsec:ondeviceperf}
In this section, we analyze the performance differences by dissecting the on-device object recognition task with an in-depth study of time breakdown, 
and resource utilization. 
We focus on understanding the performance of on-device deep inference and the implications for potential peformance improvement. 

\subsubsection{Impact of Deep Learning Models}

The choice of deep learning models including the framework and CNN model design can have a significant impact on the performance,
due to different model sizes and complexities.  
We quantify such impacts with both the model loading and probability computation time. 
We plot the loading time in Figure~\ref{subfig:model_loading_time} 
in \emph{log} scale.
For loading the same model (AlexNet and NIN),
the ported Caffe library takes up to 4.12 seconds, 
about 22X faster than using CNNDroid. Furthermore, it only takes an 
average of 103.7 ms to load the smallest SqueezeNet model.\footnote{We did not have results for running SqueezeNet using 
CNNDroid library because CNNDroid library currently does not support newer deep neural networks that include expand convolution layers, such as SqueezeNet.}
This loading happens whenever users first launch the mobile application, 
and potentially when a suspended background app is brought back. 
Our measurement of CNNDroid's long loading time suggests that users 
need to wait for up to 88 seconds to be able to interact with the mobile 
app. Although long loading time can be amortized by the number of inference 
tasks during one user interaction session, it still negatively impact user 
experiences.

Next, we show the time taken to compute the input image using 
five different configurations in Figure~\ref{subfig:model_computation_time}. 
For each configuration, we measure the computation time taken 
for all five images and collect a total of 75 data points. 
Each bar represents the average computation time across three versions 
of the same image and the standard deviation. 
CNNDroid-based AlexNet inference achieves the lowest average of 
1541.67 ms, compared to the longest time of 13745.33 ms using 
ported Caffe NIN model. Even with the fastest device-based inference, 
it still takes three times more than CPU-based cloud inference. 
In addition, we plot the end-to-end inference time in Figure~\ref{subfig:model_inference_time}. This total inference time includes the bitmap scaling 
time, the GC time, and the model computation time. 
CNNDroid-based approach takes an average of  
1648.67 ms for performing object recognition on a single image, 
about seven times faster than using ported Caffe models. 
Based on the response time rules~\cite{ws:nielsenRespTime,ws:mobileappMetrics}, 
it might lead to poor user experiences when using certain device-based inference approach.

\subsubsection{Impact of Limited Mobile Memory}
During loading CNNDroid-based models, we observe much more frequent, 
and long lasting garbage collecting activities performed by Android Runtime in our mobile device. 
When running our app using CNNDroid library, we have to 
request for a large heap of 512 MB memory.\footnote{Running 
the app with the default 192 MB memory will lead to \texttt{OutOfMemoryError}.}
Even with a large heap, the memory pressure of creating new 
objects has lead to a total of 8.33 (23.67) GC invocations when using 
CNNDroid-based AlexNet (NIN) model, as indicated in Table~\ref{tbl:gc_impact}. 
Our evaluation suggests that by allocating more memory to deep learning powered mobile apps,
or running such apps in more powerful mobile devices can mitigate the impact of garbage collection.

\subsubsection{Impact of Image Size} 
Because the CNN models only require 
images of dimension 224 by 224 pixels to perform inference tasks, 
we can scale input image to the required dimension before sending. 
Figure~\ref{fig:scaling_time} shows the time taken to scale images 
with different sizes. Each data point represents the average scaling time 
across five different runs.  The time taken to resize image grows as its size increases. 
It is only beneficial to downscale an image of size $x_1$ to $x_2$ if 
$T_{d} (x_1, x_2) + T_{n} (x_2) \leq T_{n}(x_1)$, 
where $T_{d}(x, y)$ represents the time to downscale an image from size $x$ to $y$
and $T_{n}(x)$ denotes the time to upload an image of size $x$ to a cloud server. 
For example, based on our measurement, it takes an average of 36.83 ms to upload an image of 172 KB to our cloud server. 
Also, from Figure~\ref{fig:scaling_time}, we know that it takes up to 38 ms to resize an image less than 226 KB. By combining these two observations, 
it is easy to conclude that directly uploading image one to five is more time efficient. We can expect to make informed decisions about whether 
resizing an image of size $x$ before uploading is beneficial or not given enough time measurements of resizing and uploading steps.  

\emph{\textbf{Result:} Our analysis shows that on-device inference's performance bottlenecks mainly exhibit in loading model and computing probability steps.}

\begin{figure}[t]
\centering
\begin{subfigure}{0.24\textwidth}
\includegraphics[width=\columnwidth]{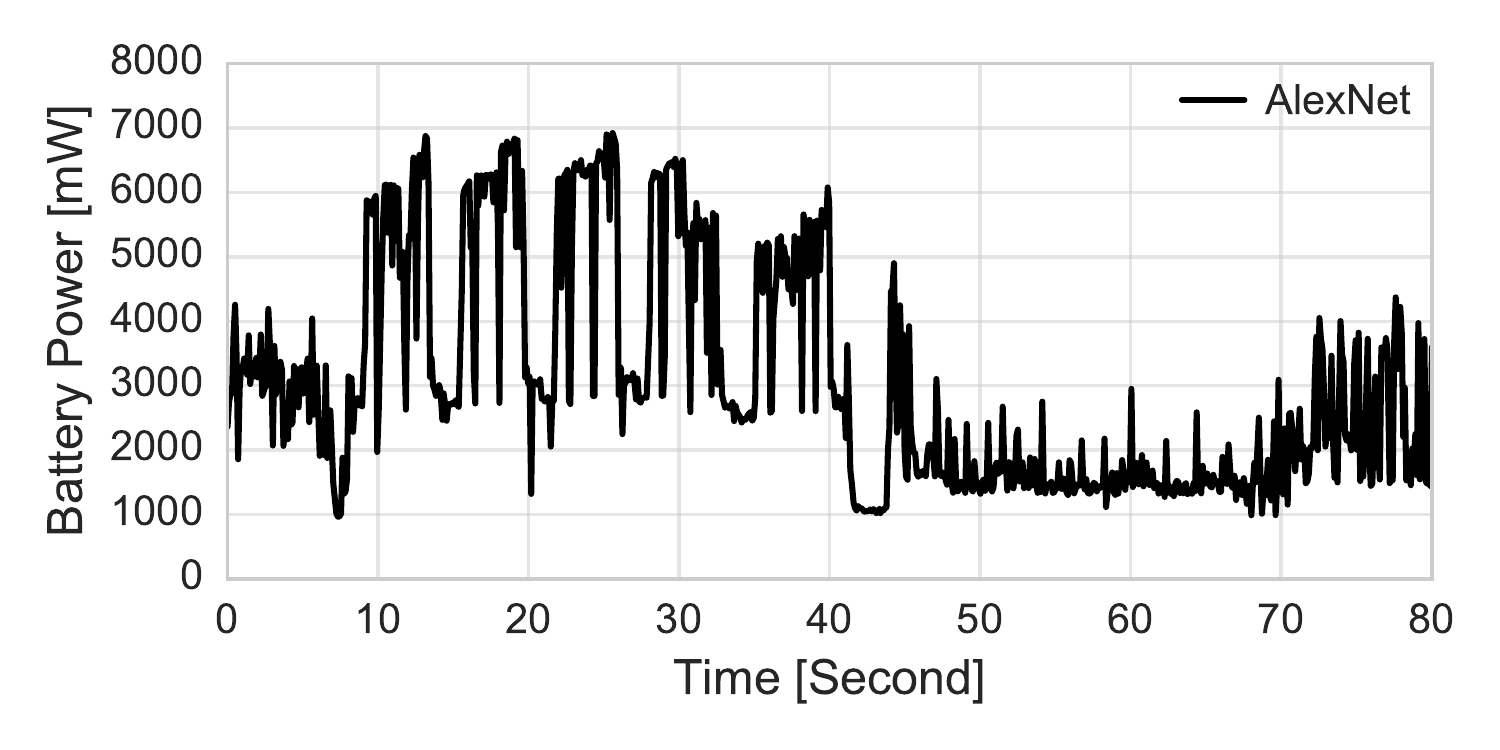}
\caption{\textnormal{Caffe-based energy.}}
\label{subfig:alexnet_battery} 
\end{subfigure}
\begin{subfigure}{0.24\textwidth}
\includegraphics[width=\columnwidth]{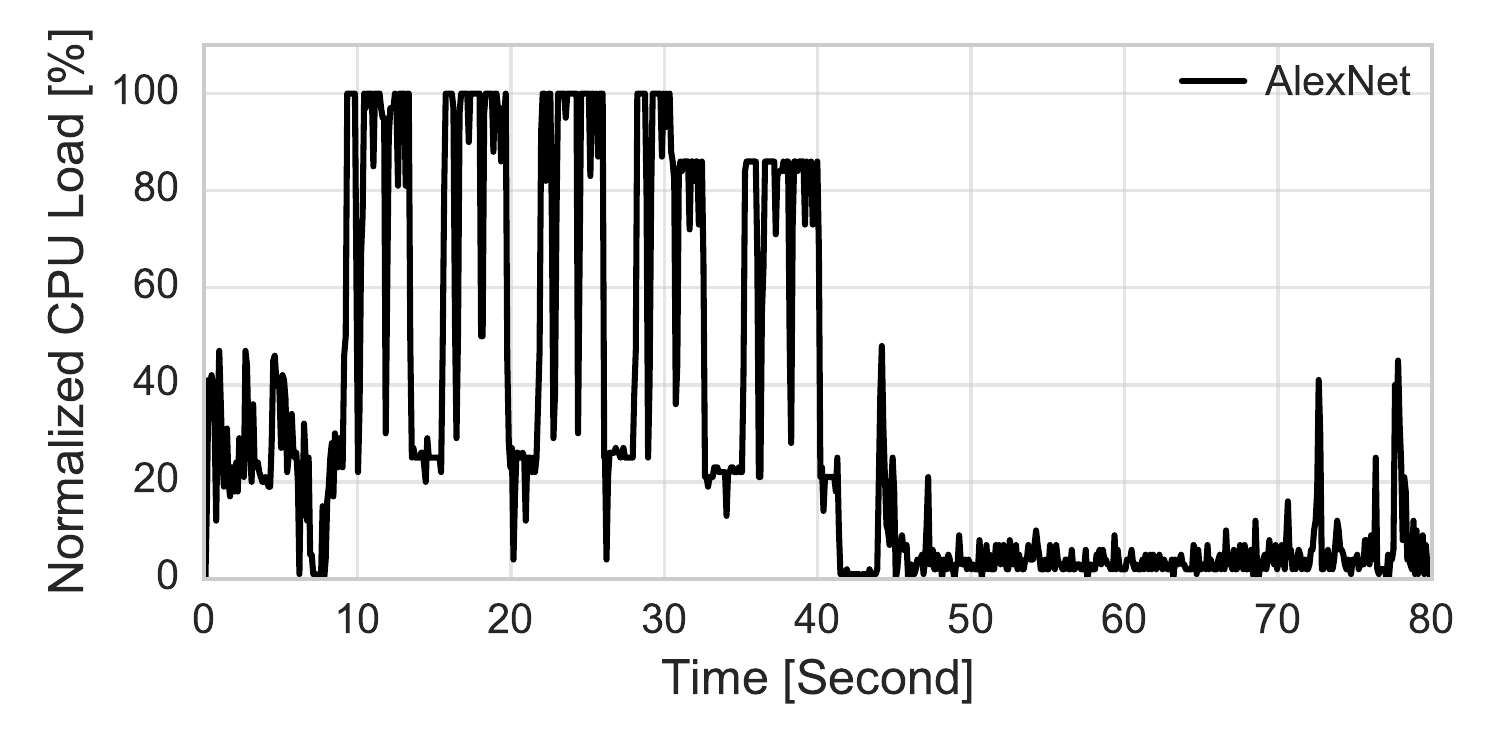}
\caption{\textnormal{Caffe-based CPU.}}
\label{subfig:alexnet_cpu_util} 
\end{subfigure}
\begin{subfigure}{0.24\textwidth}
\includegraphics[width=\columnwidth]{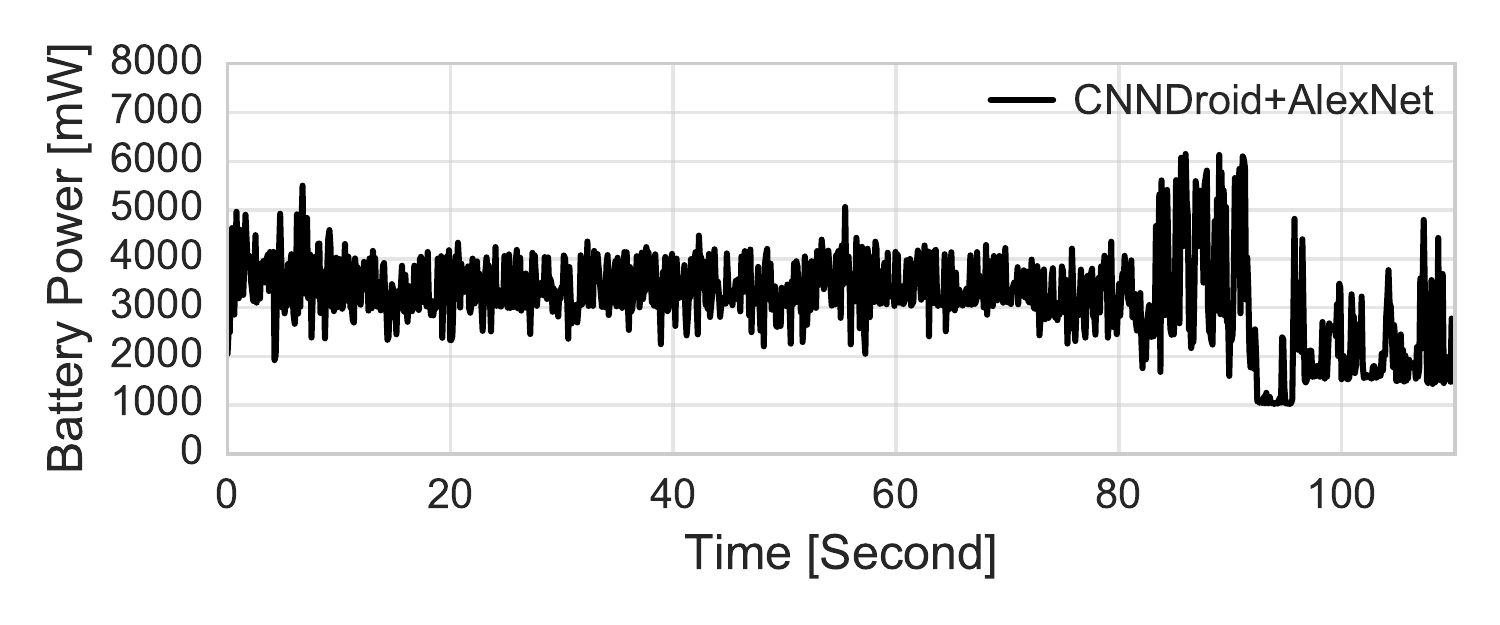}
\caption{\textnormal{CNNDroid-based energy.}}
\label{subfig:cnndroid_alexnet_energy} 
\end{subfigure}
\begin{subfigure}{0.24\textwidth}
\includegraphics[width=\columnwidth]{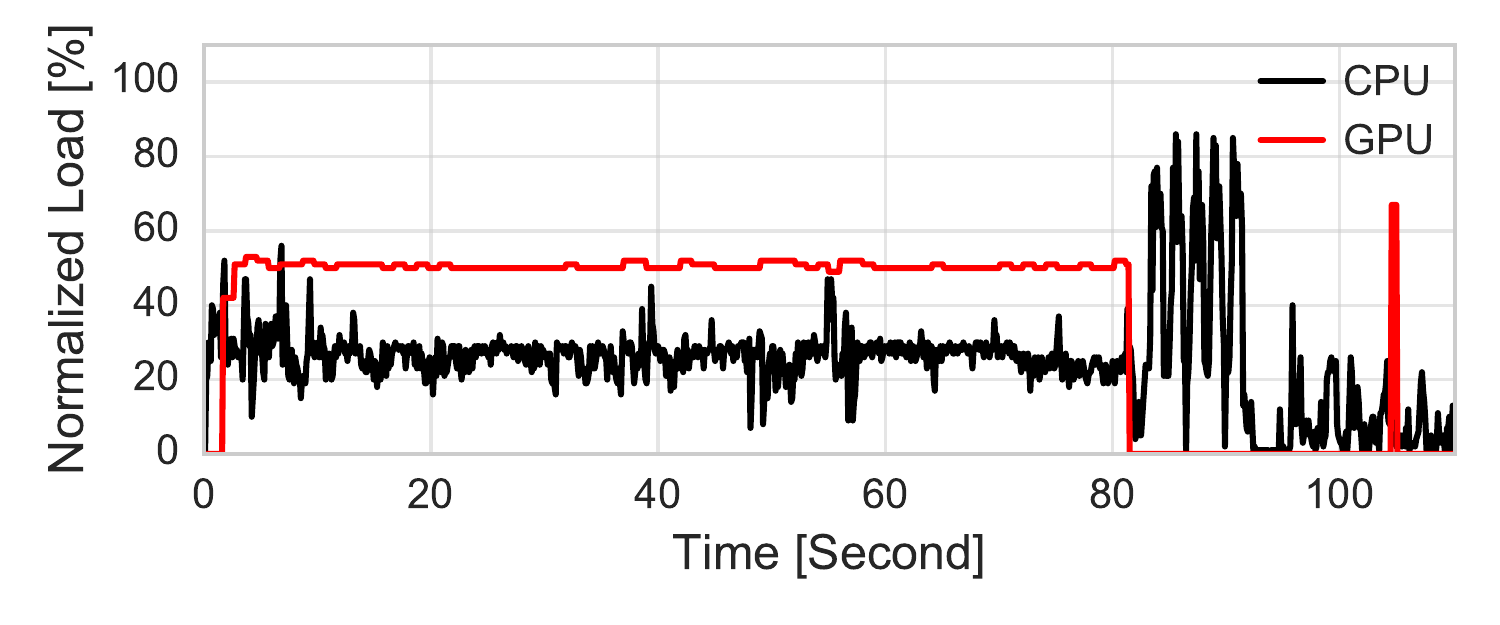}
\caption{\textnormal{CNNDroid-based resource.}}
\label{subfig:cnndroid_alexnet_cpu_gpu} 
\end{subfigure}
\caption{
Energy consumption and resource utilization of on-device object recognition task.
\todo{redo a, b to make it align better with c and d}}
\label{fig:trepn_measurement_caffe}
\end{figure}

\subsection{On-device Deep Inference Resource and Energy Analysis}

In Figure~\ref{fig:trepn_measurement_caffe}, we analyze both the energy consumption and resource utilization when running our app in different configurations \todo{using two mobile frameworks and CNN models}. We compare the time-series plots of running AlexNet model using Caffe Android library and CNNDroid framework. 
The plots correspond to experiment runs that perform inference tasks on image set one. 

For Caffe Android library based approach, we observe an initial energy consumption (and CPU utilization) that increases corresponding to loading AlexNet
CNN model into the memory, a continuation of energy spike during 
model computation, and the last phase that corresponds to displaying 
images and the most probable label texts, in Figure~\ref{subfig:alexnet_battery} and Figure~\ref{subfig:alexnet_cpu_util}. 
The other two CNN models, NIN and SqueezeNet, exhibit very similar usage pattern\footnote{Interested readers can refer to our arXiv version~\cite{guo2017mobiledl} for additional results of Caffe-based NIN and SqueezeNet models, and CNNDroid-based NIN.}. Specifically, in the case of NIN, 
the initial model loading causes the energy consumption to increase from 
baseline 1081.24 mW to up to 5000 mW;
when performing the model computation, both the energy consumption 
and CPU utilization spikes to more than 7000 mW and 66.2\%. 
Note in the case of SqueezeNet, we only observe a very small window of both energy and CPU spikes at the very beginning of measurement. This is because 
SqueezeNet can be loaded in 109 ms, compared to more than 3000 ms to load either AlexNet or NIN.

In contrast, we observe two key usage differences in CNNDroid approach, as shown in Figure~\ref{subfig:cnndroid_alexnet_energy} and Figure~\ref{subfig:cnndroid_alexnet_cpu_gpu}. First, CNNDroid-based AlexNet exhibits a longer period of more stable and lower energy consumption
compared to its counterpart in Caffe-based approach. 
This is mainly because CNNDroid explicitly expresses some of the data-parallel workload 
using \texttt{RenderScript} and is able to offload these workload to more energy-efficient mobile GPU~\cite{ws:gpuEnergy} (indicated by the high GPU utilization during model loading). 
Second, the total model computation time is significantly shortened from 40 seconds to around five seconds. In all, by shifting some of computation tasks during 
model loading, CNNDroid-based approach successfully reduces the user perceived response time. However, the CNNDroid approach consumes 85.2 mWh energy, over 42\% more than Caffe-based approach. Note 91\% of CNNDroid energy is consumed during model loading phase, and therefore can be amortized by performing inference tasks in batch.

\emph{\textbf{Result:} The CNNDroid-based approach is more energy-efficient in performing inference tasks compared to the Caffe-based approach when models are preloaded into the mobile memory.}

\section{Related Work}

To better understand the performance and power characteristics 
of modern mobile applications, researchers have developed 
a number of performance monitor tools, such as 
3GTest~\cite{Huang:2010:3gtest}, 4GTest~\cite{Huang:2012:4gtest}, AppInsight~\cite{ravindranath2012appinsight}, eprof~\cite{pathak2012energy},
and Trepn~\cite{ws:trepn} over the past decade.  
Our paper focuses on understanding the performance bottlenecks
of a new class of mobile applications that are powered by deep learning models, 
with an implemented object recognition benchmark Android app. 

In order to improve response time and preserve batter energy in resource-constrained devices, 
researchers have proposed to offload computation intensive tasks 
to the cloud~\cite{chun2011clonecloud,cuervo2010maui,mobilecomputing:reviewer,Guo2017mmsysJ}. Our paper 
confirms that cloud-based inference mobile apps still deliver better response 
time and consume less power compared to mobile-based counterparts. 
However, recent development of mobile specific deep learning optimizations~\cite{Lane:2016ko,deepsense:2016,cnndroid:2016,ws:mobilenets,ws:tensorflowlite,ws:caffe2go} and improvement in mobile GPU energy consumption~\cite{ws:gpuEnergy}
are promising improvements towards efficient mobile-based 
deep learning apps. Our work can be easily extended to evaluate the efficiency of these new models and hardware.
 \section{Conclusion} 

In this paper, we evaluate the current approaches to perform deep inference
tasks in resource-constrained mobile devices. Our analysis show that while cloud-based 
inference incurs reasonable response time and energy consumption, current 
on-device inference is only feasible for very limited scenarios.
However, with both industry and research efforts on adapting deep neural networks to the
mobile devices, we believe it is very likely that on-device inference can be done efficiently in the 
near future. 
Consequently, when
developing deep learning powered mobile apps, developers will have the freedom
to choose from cloud-based, device-based or even a hybrid approach.  

\textbf{Acknowledgements.} We thank all the reviewers for their insightful comments and Sam Ogden for proofreading, which improved the quality of this paper. 
 
\clearpage
\bibliographystyle{IEEEtran}
\bibliography{bib}
 
\end{document}